\documentclass[aps,twocolumn,superscriptaddress,showpacs,showkeys,preprintnumbers,amsmath,amssymb]{revtex4}

\usepackage{txfonts}
\usepackage{dcolumn}
\usepackage{mathrsfs}
\usepackage{bm}
\usepackage{amsmath,amssymb,epsfig,float,graphics}

\begin{document}

\title{Theoretical analysis of Casimir and thermal Casimir effect in stationary space-time}

\author{Anwei Zhang}
\email{awzhang@sjtu.edu.cn}

\address{
Department of Physics and Astronomy, Shanghai Jiao Tong University, Shanghai 200240, China
}

\begin{abstract}
We investigate Casimir effect as well as thermal Casimir effect for a pair of parallel perfectly plates placed in general
stationary space-time background. It is found that the Casimir energy is influenced by the 00-component of metric and the corresponding quantity in dragging frame. We give a scheme to renormalize thermal correction to free energy in curved space-time. It is shown that the thermal corrections to Casimir thermodynamic quantities not only depend on
the proper temperature and proper geometrical parameters of the plates, but also on the determinant of space-time metric.
\end{abstract}

\pacs{\emph{04.20.-q, 04.62.+v, 11.10.Wx}}
\keywords{ Casimir effect, thermal Casimir effect, stationary space-time}
\maketitle

The Casimir effect \cite{Casimir} is one of the most interesting
consequences of vacuum fluctuations predicted by quantum field theory.
It is originally expressed as the attraction between two neutral, perfectly conducting plates in vacuum.
In classical electrodynamics, there should be no force between neutral plates. But
the quite remarkable result actually depends on Planck's constant. Therefore, this effect is a
purely quantum effect, and results from the restriction of allowed modes in vacuum
between the boundaries.

The past few years have seen spectacular developments in Casimir effect, both theoretically
and experimentally \cite{bordag}. Space-time with nontrivial topology, is also
new element which has been taken into account \cite{dewitt,ford}.
Though no material boundaries exist, the identification conditions induced by space-time topology restrict the quantum fields modes.
Following this line, lots of investigations had been performed on the plates in non-Euclidean topology space-time \cite{Milton,Al,Bimonte}.

Recently, some authors have investigated the Casimir
effect under the influence of weak gravitational fields \cite{1,2,sorge05,sorge09,4,Bezerra142,5}. Particularly in \cite{sorge05}, the Casimir vacuum energy density between plates in a slightly curved, static space-time background was studied. Then in the weak field approximation, Bezerra et al. \cite{Bezerra142}
investigated the renormalized vacuum energy density in the plates which placed near the surface of a rotating
spherical gravitational source. Relaxing the assumption of weak field approximation, the vacuum energy in the cavity moving in a
circular equatorial orbit in the exact Kerr space-time geometry was evaluated \cite{sorge14}.
And then the thermal corrections in such a case were calculated \cite{an}.

The main purpose of this article is to generalize the results in \cite{sorge05,Bezerra142,sorge14,an} to a more general case: stationary space-time.
We will theoretically analyze the general properties of Casimir energy as well as thermal corrections for cavity in such a general space-time. Besides, we will give another renormalization scheme for Casimir thermal corrections in curved space-time.

We start by defining a local Cartesian coordinate frame ($x, y, z$) attached to a pair of parallel perfectly conducting
plates separated by a distance $L$, with $z$ axis being perpendicular to the plates and the origin located at the center of the
apparatus. In such a local frame, the general
stationary space-time background metric can be written as
\begin{equation}\label{1}
  ds^2=g_{00}(z)dt^2+g_{11}(z)dx^{2}+g_{22}(z)dy^{2}+g_{33}(z)dz^{2}+2g_{03}(z)dtdz.
\end{equation}
As a preliminary attempt, we only consider the case that the metric is dependent on coordinate $z$. It can be seen that
 this metric possesses $\partial_{t}, \partial_{x}, \partial_{y}$ as killing vector, so the massless scalar field confined in the
 plates has the form \cite{ss}
 \begin{equation}\label{2}
   \phi_n=N_n \textrm{exp}(-i\omega_n t+ik_{x}x+ik_{y}y)\sin\bigg(\frac{n \pi}{L}z\bigg)f(z),
 \end{equation}
where $N_n$ is a normalization constant, the sine function stems from standing-wave condition and Dirichlet boundary condition which requires the $\phi_n$ to be zero at the boundaries of the plates, and $f(z)$ is an unknown function of $z$ which can be solved by using Klein-Gordon equation
\begin{equation}\label{3}
 \frac{1}{\sqrt{-{g}(z)}}\partial_\mu[\sqrt{-{g}(z)}{g}^{\mu\nu}(z)\partial_\nu]\phi_n=0
\end{equation}
with $g(z)=\textrm{det}(g_{\mu\nu}(z))$. After some calculations, we can have
$f(z)= \textrm{exp}(i\omega_n g^{03}/g^{33})$ and
\begin{equation}\label{4}
 \omega_n=\sqrt{\frac{g^{33^2}}{g^{03^2}-g^{00}g^{33}}}\sqrt{\frac{g^{11}}{g^{33}}k^{2}_{x}+\frac{g^{22}}{g^{33}}k^{2}_{y}+\bigg(\frac{n\pi}{L}\bigg)^{2}}.
\end{equation}
Note that for simplicity, we have taken the approximation $\int_{-L/2}^{L/2}k_zdz\approx k_zL$ in standing-wave condition and $\partial_{z}(\sqrt{-g(z)}g^{3\nu}(z))\approx0$ in Klein-Gordon equation. These approximations can be fulfilled when
the metric is independent of coordinate $z$ or the distance $L$ between the plates is very small \cite{sorge14} so that we only need expand $g_{\mu\nu}(z)$ to zero order $g_{\mu\nu}(0)$ \cite{qw}. Actually the separation between plates is on atomic or subatomic scale and when $L=1\mu m$, the plates is separated relatively large \cite{bordag}. Thus the
 above calculations are valid in zero-order approximation for realistic plates in the curved space-time background (\ref{1}).

The parameter $N_n$ can be obtained from the scalar product \cite{Birrell}
\begin{equation}\label{4}
  (\phi_n, \phi_m)=i\int_\Sigma[(\partial_\mu\phi_n)\phi^{\ast}_m-\phi_n(\partial_\mu\phi^{\ast}_m)]\sqrt{g_s}n^{\mu}d\Sigma,
\end{equation}
where $g_s=-g/g_{00}$ is the
determinant of the induced metric on space-like hypersurface $\Sigma$, $n^{\mu}$ is a time-like unit vector and in our case it is $(\sqrt{g^{00}},0,0,-g_{03}\sqrt{g^{00}}/g_{33})$. Then from orthogonality condition, we arrive at
\begin{equation}\label{5}
 N^{2}_n=\frac{g_{00}\sqrt{-g_{00}g_{11}g_{22}g_{33}}}{-g(2\pi)^{2}L\omega_n}.
\end{equation}
Here we have used the following relations:
 \begin{eqnarray}
  -g &=& g_{11}g_{22}(g^{2}_{03}-g_{00}g_{33}), \,\,g^{00}=\frac{-g_{33}}{g^{2}_{03}-g_{00}g_{33}},\,\,g^{11}= \frac{1}{g_{11}},\nonumber\\
  g^{22}&=&\frac{1}{g_{22}},  \,\, g^{33}=\frac{-g_{00}}{g^{2}_{03}-g_{00}g_{33}},\,\,g^{03} =\frac{g_{03}}{g^{2}_{03}-g_{00}g_{33}}.
 \end{eqnarray}

Now we proceed to investigate Casimir energy in the cavity which should take the form
\begin{equation}\label{6}
  E_0=\int_V\langle 0|T_{\mu\nu}|0\rangle U^{\mu}U^{\nu}\sqrt{g_s}dV,
\end{equation}
in which $ U^{\mu}$ is the 4-velocity of observer and it is $(1/\sqrt{g_{00}},0,0,0)$ for static observer located at coordinate origin, $\langle 0|T_{\mu\nu}|0\rangle$ is the expected value of energy-momentum tensor and its $00$-component reads
\begin{equation}\label{7}
 \langle 0|T_{00}|0\rangle=\sum_n\int\int(\partial_t \phi_n\partial_t\phi^{\ast}_n-\frac{1}{2}g_{00}g^{\mu\nu}\partial_\mu\phi_n\partial_\nu\phi^{\ast}_n)dk_xdk_y.
\end{equation}
 Performing the integral in $z$ in (\ref{6}), taking necessary variable substitution and introducing an exponential cutoff function so as to renormalize the divergent energy, finally we can obtain the Casimir energy observed at coordinate origin
\begin{equation}\label{8}
  E_0=\sqrt{\frac{g_{00}}{\hat{g}_{00}}}E_p,
\end{equation}
where $\hat{g}_{00}=g_{00}-\frac{g^{2}_{03}}{g_{33}}$ is the 00-component of metric in dragging frame, $E_p=-\frac{\pi^{2}S_p}{1440L^{3}_p}$, $S_p=\int\int\sqrt{g_{11}g_{22}}dxdy$ and $ L_p=\int^{L/2}_{-L/2}\sqrt{-g_{33}+\frac{g_{03}^{2}}{g_{00}}}dz$ denote proper surface area and proper length of the cavity, respectively, so $E_p$ is the Casimir energy in flat space-time.

From the expression (\ref{8}), one can see that for comoving observer in static space-time background, the  Casimir energy is just the proper value in Minkowski space-time. This basically agrees with the result obtained for the cavity placed in weak gravitational field \cite{sorge05,sorge09,Bezerra142}. One can also find that when the observer is in stationary rather than static space-time, the observed value will be different with the proper value
by a factor depending on space-time background. It can be checked that the result in \cite{sorge14} is only a particular case of (\ref{8}). The above conclusions can be understood by taking a Coordinate scale transformation $\sqrt{|g_{\mu\mu}|}dx^{\mu}=dx'^{\mu}$, then line element becomes $\eta_{\mu\mu}dx'^{\mu}dx'^{\mu}+2\frac{g_{03}}{\sqrt{-g_{00}g_{33}}}dt'dz'$ with $\eta_{\mu\mu}$ being Minkowski metric. In static case that $g_{03}=0$, the space-time is actually flat in the rescaled Coordinate, thus we can have $E_0=E_p$.
But in the case of $g_{03}\neq0$, the space-time is still curved after transformation, so generally $ E_0\neq E_p$. We should note that such a Coordinate transformation does not change the Casimir energy which can be easily verified. The above discussion is only limited to comoving observer case.
Actually, the observed Casimir energy is dependent on observer. For an arbitrary stationary observer located at point $z$, this energy should be
\begin{equation}\label{9}
   E_z=\frac{g_{00}}{g_{00}(z)}\sqrt{\frac{g_{00}}{\hat{g}_{00}}}E_p.
\end{equation}
The above equation can be rewritten as:
\begin{equation}\label{90}
g_{00}(z)E_z=g_{00}E_0=const.
\end{equation}
Thus the Casimir energy observed at one point is inversely proportional to the 00-component of metric at this point, regardless of the space-time is static or stationary.

In practice, Casimir cavity is immersed in thermal bath. So in the next we will take temperature into account and give thermal corrections to the Casimir thermodynamic quantities. We begin with the thermal correction to free energy \cite{mal,an}
\begin{equation}\label{11}
 \Delta_TF=\frac{1}{L}\sum^{\infty}_{n=1}\int^{\infty}_{-\infty}
\frac{dk_xdk_y}{(2\pi)^{2}}\int_V dxdydz\sqrt{g_s}\; T\ln\big(1
-e^{-\omega_n/T}\big).
\end{equation}
For the convenience of calculation, here we assume the temperature is independent of coordinate, so it can be extracted from the integral.
Now substituting (\ref{4}) in (\ref{11}) and taking $\tilde{k}_x=\sqrt{\frac{g^{11}}{g^{33}}}k_x$, $\tilde{k}_y=\sqrt{\frac{g^{22}}{g^{33}}}k_y$, $\beta=\sqrt{\frac{g^{33^2}}{g^{03^2}-g^{00}g^{33}}}/T$, then after some algebra, we arrive at
\begin{equation}\label{12}
 \Delta_TF=\frac{-g^{33}\sqrt{g_{00}}S_p}{2\pi \sqrt{g^{11}g^{22}}\beta }\sum^{\infty}_{n=1}\int^{\infty}_{0}kdk\ln\big(1
-e^{-\beta\sqrt{k^{2}+(\frac{n\pi}{L})^{2}}}\big)
\end{equation}
with $k^{2}=\tilde{k}^{2}_x+\tilde{k}^{2}_y$.  The logarithm in (\ref{12}) can be written as power series
\begin{equation}\label{13}
\Delta_TF=\frac{g^{33}\sqrt{g_{00}}S_p}{2\pi \sqrt{g^{11}g^{22}}\beta^{3} }\sum^{\infty}_{n,m=1}\frac{1}{m}\int^{\infty}_{n\pi \beta/L}hdh
e^{-mh},
\end{equation}
where $h=\beta\sqrt{k^{2}+(\frac{n\pi}{L})^{2}}$. Now performing the integral in (\ref{13}) and then taking the summation in $n$, one can get
\begin{eqnarray}\label{14}
 \Delta_TF&=&-\frac{\sqrt{-g}S_p}{32\pi L^{3}_p}\sum^{\infty}_{m=1}\bigg[\frac{\coth(\pi m\tilde{\beta})}{(m\tilde{\beta})^{3}}
+\frac{\pi}{(m\tilde{\beta})^{2}\sinh^{2}(\pi m\tilde{\beta})}\bigg]\nonumber\\
&&+\frac{\zeta(3)\sqrt{-g}S_p}{32\pi (L_p\tilde{\beta})^{3}},
\end{eqnarray}
where $\tilde{\beta}=\beta/2L=1/(2T_pL_p)$ is a dimensionless parameter, $T_p=T/\sqrt{g_{00}}$ is the proper temperature and $\zeta(3)=\sum^{\infty}_{m=1}1/m^{3}$ is the Riemann zeta function.

It can be seen from (\ref{14}) that the quantity $\Delta_TF/\sqrt{-g}$ is only dependent on the proper geometrical parameters of the cavity and proper temperature. So to renormalize such a quantity, we can take the cavity out of the stationary space-time, put it in flat background and let the proper temperature unchanged. Thus such a quantity will be still itself. Now let us analyze the asymptotic behave of $\Delta_TF/\sqrt{-g}$ under the condition of $L_p\rightarrow\infty$, which corresponds to $\tilde{\beta}\ll1$. This can be done by performing series expansion of this quantity in the neighborhood of $\tilde{\beta}=0$, and then taking summation in $m$. The result is
\begin{eqnarray}\label{15}
-V_p\frac{\pi^{2}T^{4}_p}{90}+S_p\frac{\zeta(3)T^{3}_p}{4\pi}
-\frac{\pi^{2}S_p}{720L^{3}_p}.
\end{eqnarray}
Here we have used the properties $\zeta(4)=\pi^{4}/90$ and $\zeta(-n)=-B_{n+1}/(n+1)=0$ for even integers $n$,
$B_{n+1}$ is a Bernoulli number and $V_p=S_pL_p$ is proper volume of the cavity. The last term in (\ref{15}) can be omitted since it is already zero in the limit condition. We know that thermal correction to free energy should be null in the limit case $L_p\rightarrow\infty$, so the needed renormalized value can be obtained by subtracting the terms in (\ref{15}), thus we have
\begin{eqnarray}\label{16}
\Delta_TF/\sqrt{-g}&=&-\frac{\hbar cS_p}{32\pi L^{3}_p}\bigg\{\sum^{\infty}_{m=1}\bigg[\frac{\coth(\pi m\tilde{\beta})}{(m\tilde{\beta})^{3}}
+\frac{\pi}{(m\tilde{\beta})^{2}\sinh^{2}(\pi m\tilde{\beta})}\bigg]\nonumber\\
&&-\frac{\pi^{3}}{45\tilde{\beta}^{4}}\bigg\}=\Delta_TF_p,
\end{eqnarray}
where $\Delta_TF_p$ denotes the proper thermal correction to free energy. To show this proper thermal correction clearly, we illustrate it in Fig. 1. Here we introduce a truncation $m=500$, since in the case of lager $m$, the value in the square bracket of (\ref{16}) tends to zero. It can be seen that at high temperature, the proper thermal correction to free energy is proportional to the proper temperature \cite{bordag}.
 \begin{figure}[htbp]
\centering
\includegraphics[width=0.457\textwidth ]{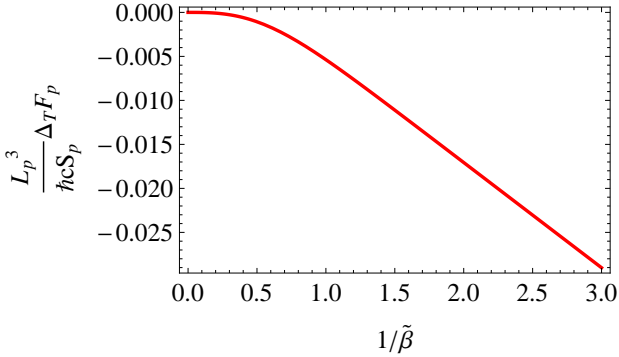}
\caption{ (color online).  $\frac{L^{3}_p}{\hbar cS_p}\Delta_T F_p$ as a function of $1/\tilde{\beta}=2k_BT_pL_P/\hbar c$.
}\label{Fig.1}
\end{figure}

It is worthwhile to note that there is another renormalization approach \cite{mal} via subtracting the terms $\alpha_0T^{4}, \alpha_1T^{3}$  and $\alpha_2T^{2}$ which are contained in the asymptotic high temperature limit of the free energy. Here the coefficients $\alpha_0, \alpha_1, \alpha_2 $ have
something to do with the geometrical parameters of the plates \cite{bordag}. This method has been widely used including in the case of the apparatus placed in the curved background \cite{Bezerra11,Bezerra14}.  Although the physical mechanism of this approach seems to be obscure, it gives the same result as ours $(\ref{16})$.

Now the total renormalized Casimir free energy can be written as
\begin{equation}\label{17}
 F=E_z+\sqrt{-g}\Delta_TF_p,
\end{equation}
in which the term $-\sqrt{-g}\frac{\hbar cS_p}{32\pi L^{3}_p}\frac{\pi^{3}}{45\tilde{\beta}^{4}}=-\sqrt{-g}V_p\pi^{2}T^{4}_p/90$ is actually the free energy of black body radiation. This can be verified by calculating the formula for free energy density of black body radiation
$T\int^{\infty}_{-\infty}
\frac{dk_xdk_ydk_z}{(2\pi)^{3}}\ln\big(1
-e^{-\omega_n/T}\big)$.
Based on (\ref{17}), we can deduce other thermodynamic quantities, such as Casimir internal energy. If such a quantity is defined by $U\equiv-T^{2}_p\frac{\partial(F/T_p)}{\partial T_p}$, then we can have $U=E_z+\sqrt{-g}\Delta_TU_p$ with $\Delta_TU_p=-T^{2}_p\frac{\partial(\Delta_TF_p/T_p)}{\partial T_p}$. We can also define Casimir entropy as $S\equiv-\frac{\partial F}{\partial T_p}=\sqrt{-g}\Delta_TS_p$ with $\Delta_TS_p=\frac{\partial\Delta_TF_p}{\partial T_p}$ and Casimir heat capacity at constant volume $C_{V_p}\equiv(\frac{\partial U}{\partial T_p})_{V_p}=\sqrt{-g}\Delta_TC_{V_p}$ with $\Delta_TC_{V_p}=\frac{\partial\Delta_TU_p}{\partial T_p}$. From Fig. 2, it can be found that (i) as the proper temperature approaches to zero, the Casimir internal energy and entropy  respectively go to $E_z$ and $0$ which is compatible with the third law of thermodynamics; (ii) at high temperature limit, the Casimir internal energy and entropy tend to constant values, thus they are independent of the proper temperature in this condition; (iii)  the Casimir heat
capacity at constant volume has a maximum when $1/\tilde{\beta}$ approximately reaches to  $0.628$, at which point Casimir internal energy
changes fastest.
 \begin{figure}[htbp]
\centering
\includegraphics[width=0.457\textwidth ]{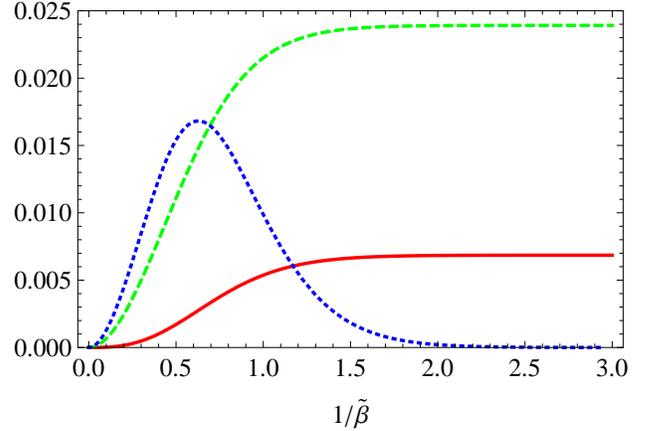}
\caption{(color online).   $\frac{L^{3}_p}{\hbar cS_p}\Delta_T U_p$ (solid line), $\frac{L^{2}_p}{k_BS_p}\Delta_T S_p$ (dashed line) and
 $\frac{L^{2}_p}{k_BS_p}\Delta_TC_{V_p}$ (dotted line)
as a function of $1/\tilde{\beta}$, respcetively.
}\label{Fig.2}
\end{figure}

From the expressions of thermodynamic quantities, it can be seen that the thermal corrections to the thermodynamic quantities between plates, which are placed in general stationary space-time, generally depend on not only the proper temperature and the proper geometrical
parameters of the cavity, but also the background space-time. This is different with the result of \cite{an}, in which the thermal corrections have nothing to do with the determinant of metric. This is due to that Kerr space-time is a very special case and when the plates are placed in such a space-time, the the determinant of metric between the plates can be $1$. So in such a case, the thermal corrections are independent of background space-time if a single
reservoir is located in the plates.

In summary, we have investigated Casimir energy and the thermal corrections for cavity in general stationary space-time background. It is shown that the Casimir energy is dependent on the ratio between 00-component of metric and the corresponding value in dragging frame. Besides, the Casimir energy observed is inversely proportional to the 00-component of metric at observation point.
We give a renormalization scheme for Casimir thermal correction in curved space-time and explore the general properties of Casimir free energy, internal energy, entropy as well as heat capacity at constant volume. It is found that the thermal corrections to these thermodynamic quantities not only depend on the proper temperature and the proper geometrical
parameters of the cavity, but also on the determinant of metric. This result may provide a clue to advance the field of relativistic thermodynamics.

We would like to thank Q. Huang and Y. Jin for some help.


\begin{thebibliography}{99}
\bibitem{ss} Throughout the text, the natural units will be used.
\bibitem{qw} In this letter $g_{\mu\nu}(0)$, $g^{\mu\nu}(0)$, $g(0)$ are abbreviated to $g_{\mu\nu}$, $g^{\mu\nu}$ and $g$, respectively.
\end{thebibliography}
\end{document}